\documentclass[12pt,letterpaper]{article}
\usepackage{epsfig}
\usepackage{epstopdf}
\usepackage[utf8]{inputenc}
\usepackage{amsfonts}
\usepackage{amsmath}
\usepackage{amssymb}
\usepackage{amsfonts}
\usepackage{graphicx}
\usepackage{amsthm}
\usepackage{epsfig}
\usepackage{color}
\usepackage{epstopdf}
\usepackage{setspace}
\usepackage{tocbibind}
\usepackage{lipsum}
\usepackage{textcomp}
\usepackage{longtable}
\usepackage{verbatimbox}
\usepackage{array}
\usepackage{mathtools}
\usepackage{caption}
\usepackage{amsmath,amssymb,amsfonts}
\usepackage{algorithmic}
\usepackage{graphicx}
\usepackage{textcomp}
\usepackage{wrapfig}
\usepackage{mathtools, cuted}
\begin{document}
\huge{Phase shift-cavity ring down spectroscopy in linear and active fiber cavities for sensing applications at 1550 nm}\\

\normalsize{ Ubaid Ullah and M. Imran Cheema $^{*}$} \\

$^1$Department of Electrical Engineering, Syed Babar Ali School of Science and Engineering, Lahore University of Management Sciences, Lahore 54792 Pakistan\\

$^*$imran.cheema@lums.edu.pk\\ 

\section{Abstract}
Liquid phase sensing applications at 1550~nm are highly desirable due to widely available off-the-shelf components. Generally, liquids at 1550~nm induce a high absorption loss that limits the overall sensor's sensitivity and detection limit. One solution is to use an active fiber loop in conjunction with cavity ring down spectroscopy to overcome these absorption losses. However, the amplifier inside the fiber loop suffers from inherent gain fluctuations that limit the sensing system's overall performance. Here, we provide a novel sensor using the wavelength-scanned phase shift-cavity ring down spectroscopy (PS-CRDS) in conjunction with a linear active fiber cavity that potentially offers a more sensitive solution than traditional fiber loop sensors. We use a tapered fiber as a sensing head inside the active cavity built from fiber Bragg gratings. We derive a theoretical phase shift expression for our system and simulate it using the finite element method to determine optimum tapered fiber diameter for glucose sensing in DI water. Compared to a non-amplified system, we find that our amplified system can increase the sensitivity by fourteen times via the amplifier gain tuning. We also conduct experimental measurements using 0-15.5~mM glucose solutions and find them in excellent agreement with our theoretical predictions. Experimentally we obtain the sensor's sensitivity of 0.768~$^o$/mM (1164~$^o$/RIU) and detection limit of 0.75~mM ( 4.5~$\times$~10$^{-4}$~RIU) without any temperature stabilization in the system. We anticipate that the present work will find a wide range of sensing applications in fiber cavities, ring resonators, and other microcavity structures.
\section{Introduction}
\label{sec:introduction}
Phase shift-cavity ring down spectroscopy (PS-CRDS), also called as cavity attenuated phase shift spectroscopy (CAPS) is a multipass sensitive technique independent of the source intensity fluctuations. In PS-CRDS, a CW laser source is sinusoidally modulated at an angular frequency, $\omega$,  and is injected into a cavity. The output of the cavity has a phase-shift, $\phi$,  compared to the input. The phase shift is directly related to the cavity ring down time, $\tau$, as, $tan\phi =-\omega\tau$ \cite{herbelin1981, berden2009cavity}. The change in the ring down time as a function of losses in a sample filled cavity can then be utilized for various sensing applications. For example, PS-CRDS has been utilized in numerous sensing applications for detecting gases and particles, including nitrogen dioxide \cite{kebabian2008practical}, oxygen \cite{engeln1996phase}, ethene \cite{grilli2010phase}, acetylene \cite{hovde2015phase}, peroxy, organic nitrates \cite{sadanaga2016thermal}, and soot particles ejecting from $V2527$ aircraft engine \cite{yu2011direct}. All of these applications utilize free space cavities for detection purposes.

Being a highly sensitive technique, PS-CRDS is highly attractive for chemical identification and biosensing applications. However, free space cavities are not suitable for liquid phase applications, primarily for a couple of reasons. One, the cavity becomes highly unstable due to the misalignment of optics, contamination, and the liquid inside the cavity. Second, the cavity mirrors degrade over time due to continuous contact with the liquid \cite{vallance2014cavity, thompson2017cavity}. To minimize the impact of these issues, researchers have looked over various cavity designs in terms of Brewster cavities, thin sample cells, and liquid jets inside the cavity \cite{Rushworth_2012}. Nevertheless, the cavity instability problem is not entirely solved, and the overall designs are also highly complicated for developing a viable liquid phase PS-CRDS sensor with free space cavities \cite{thompson2017cavity}.

The problems mentioned above for liquid phase sensing applications can be solved by using optical fiber loops. Passive fiber loops in conjunction with the conventional cavity ring down spectroscopy (CRDS) and PS-CRDS have been used for various liquid sensing applications \cite{waechter2009405, li2006capillary, barnes2008chemical}. A passive loop generally contains a tapered fiber \cite{sharma2017fiber} or microgap \cite{Tong_2004} as a sensing head. The passive loops are typically several tens of meters and require weak couplers for coupling light into and out of the loop.

Liquid phase sensing applications at 1550~nm are highly attractive as various cost-effective, and ready-made optical components are available to build a viable optical sensor. One of the major problems with passive loops towards biosensing or chemical identification applications is that they produce a high background absorption for aqueous solutions at 1550~nm. This problem can be minimized by using an EDFA amplifier in the fiber loop \cite{Stewart_2001}. CRDS in conjunction with active fiber loops have been then utilized for various gas and liquid phase applications \cite{Stewart_2004,Liu_2010,sharma2018comparison}. An amplifier in the fiber loop minimizes the background absorption and increases the overall sensitivity and detection limit of the sensing system.

Although CRDS in an active loop improves on background liquid absorption at 1550 nm, however, gain fluctuations and nonlinearity of the amplifier seriously impacts the CRDS measurements \cite{chu2019modeling, chu2020modeling}. It happens because in CRDS, after every round trip, the pulse strength decreases, and as a result, the amplifier response varies with varying input powers. This process distorts the exponential decay curve of the ring down time, and hence noise is added in the overall measurement.

\begin{figure}[htbp!]
\centering
\includegraphics[width=8cm]{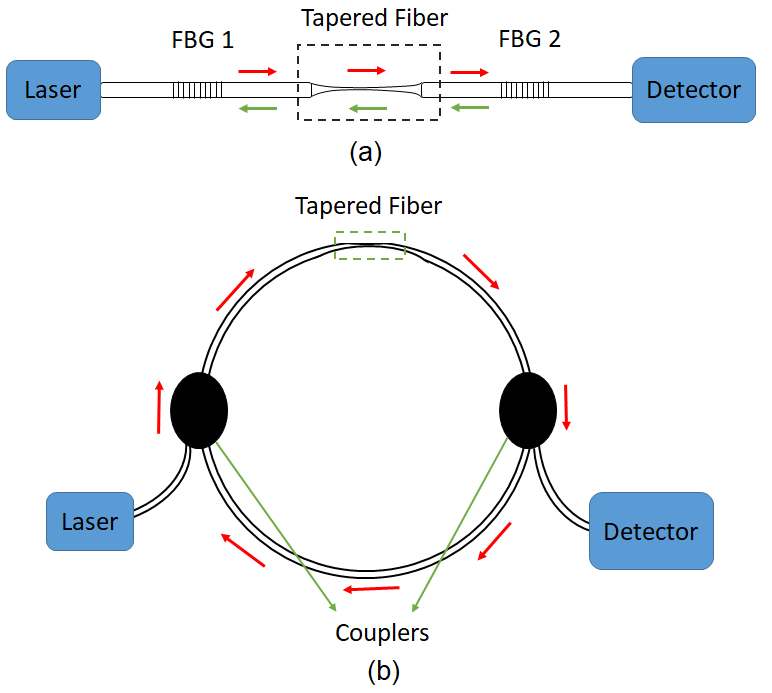}
\caption{In one round trip, the propagating optical field in linear cavity (Fig. (a)) samples the analyte twice  but the loop cavity (Fig. (b)) samples it only once.  FBG- Fiber Bragg grating. }
\label{fig:loopvslinear}
\end{figure}

Compared to fiber loops, a linear cavity is more sensitive for sensing measurements as the circulating optical field samples an analyte twice in one round trip, as also clearly shown in Fig. \ref{fig:loopvslinear}. This suggests that if we use PS-CRDS in linear active cavities, we can not only have a more sensitive approach at 1550 nm but we can avoid the amplifier’s impact on the circulating signal. This is because in PS-CRDS, not only the sinusoidal signal amplitude remains constant for all times, but we also primarily measure the phase difference between the input and output sinusoid and not the signal strength. Therefore, the present work gives the first and novel scheme of an active linear cavity at 1550 nm in conjunction with PS-CRDS measurements for liquid phase sensing applications.

We build a novel configuration of an active linear fiber cavity in the current work using circulators, a tapered fiber, a semiconductor optical amplifier, and fiber Bragg gratings. We apply the wavelength-scanned PS-CRDS approach \cite{cheema2012simultaneous, ghauri2021detection} to the built amplified cavity to carry out sensing measurements with various aqueous glucose concentrations. We derive the theoretical PS-CRDS phase expression for the sensing configuration and also determine an optimum tapered fiber diameter to maximize the sensor's sensitivity towards refractometric liquid sensing applications. In literature, various demonstrated evanescent field sensors use either absorption \cite{tarsa2004evanescent} or refractive index change \cite{ni2008fiber} of an analyte in response to a sensing event. However, a complete model will involve both the absorption and refractive index changes of the analyte. Here, we consider the impact of both the absorption and refractive index changes of the liquid surrounding the tapered fiber on the output PS-CRDS signal. We show that the sensitivity can be increased by fourteen times compared to the non-amplified one by using the amplified linear cavity. We experimentally measure glucose solution concentration employing a tapered fiber of length $\approx$ 2.5 mm and diameter $\approx$ 7 $\mu m$. We find that our setup offers the sensitivity of 0.768$^o$/mM (1164$^o$/RIU) and detection limit of 0.75 mM ( 4.5 $\times$ 10$^{-4}$ RIU) for a 0-15.5~mM glucose concentration range. These numbers are achieved without any temperature controllers in the fluidic cell and fiber Bragg gratings.

We now describe the rest of the paper. In Section \ref{sec:Theoretical Formulation}, we provide theoretical PS-CRDS phase expression and the sensing principle followed by simulation results on the sensor's sensitivity in Section \ref{sec:Sim}. We give experimental results in Section \ref{sec:Exp} and finally we conclude our work in Section \ref{sec:concl}.
\section{Theoretical Formulation}
\label{sec:Theoretical Formulation}
\begin{figure*}[ht]
\centering
\includegraphics[width=\linewidth]{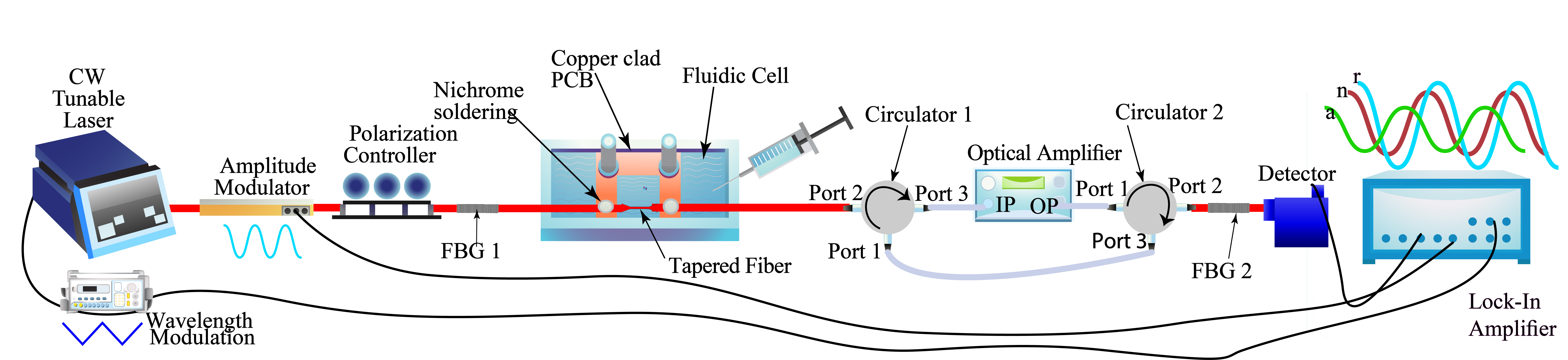}
\caption{Experimental Schematics. With respect to the reference sinusoid (`r'), PS-CRDS phase of the amplified signal (`a') is greater than the non-amplified (`n') one. FBG- Fiber Bragg grating.}
\label{Ch4_Fig_1}
\end{figure*}

The schematics of the proposed active linear cavity with PS-CRDS measurements are shown in Fig.~\ref{Ch4_Fig_1}.  The CW laser’s current is modulated by a triangular wave to linearly sweep the laser wavelength, ensuring that the cavity resonates periodically.  An external Mach-Zehnder modulator is employed to sinusoidally modulate the laser with a frequency of $\omega_m$ for the PS-CRDS measurements. The modulated light enters into the optical cavity through FBG~1. The light propagates through the tapered fiber (sensing head) that induces loss corresponding to an analyte under test. The forward light encounters the first circulator that guides light towards a silicon optical amplifier through port $2\rightarrow 3$. The amplified signal completes half of the round trip after passing through port $1\rightarrow 2$ of the second circulator and reaches FBG~2. A small portion of the light is transmitted while a large portion gets reflected. The reflected light passes through port $2\rightarrow 3$ of circulator~2, then through port $1\rightarrow 2$ of circulator~1, transverses tapered fiber, and finally reaches FBG~1, where it completes one round trip. Upon reflection from FBG~1, the second round trip starts. The transmitted light enters into the lock-in amplifier via a detector, and hence the phase shift between cavity transmission and reference modulation signal is measured. The cavity builds up at the resonance and generates a maximum phase delay measured as a phase dip around a transmission signal maximum via the lock-in amplifier. On increasing the amplifier gain, losses decrease in the cavity, and thus the phase shift increases. Note that the circulators are unidirectional, i.e., no signal can go from port 3 to port 2.

The overall transmission field is the superposition of several transmission terms, which denote the throughput of the cavity after successive round trips. After several algebraic steps on similar lines as done in \cite{yariv2007photonics} for standard Fabry-P$\acute{\text{e}}$rot cavities, we obtain the following transmission expression for the overall cavity setup shown in Fig. \ref{Ch4_Fig_1}:
\begin{strip}
\begin{equation}
\label{Cha4_Eq:9}
T = \frac{L_{cn}^5 L_t L_{c1}^{2\rightarrow 3} L_{c2}^{1\rightarrow 2} G (1-R_1)(1-R_2)}{\bigg[\bigg( 1-\sqrt{L_t^2 L_{cn}^9 L_{c1}^{1\rightarrow 2} L_{c1}^{2\rightarrow 3} L_{c2}^{1\rightarrow 2} L_{c2}^{2\rightarrow 3} G R_1 R_2} \bigg)^2 \\ + 4 \sqrt{L_t^2 L_{cn}^9 L_{c1}^{1\rightarrow 2} L_{c1}^{2\rightarrow 3} L_{c2}^{1\rightarrow 2} L_{c2}^{2\rightarrow 3} G} \sqrt{R_1 R_2} \sin^2 ((\delta_f + \delta_b)/2)\bigg]},
\end{equation}
\end{strip}
where $L_{cn}$, $L_t$, $L_{c1}$, and $L_{c2}$ represent losses due to fiber connectors, tapered fiber, circulator~1, and circulator~2, respectively. The amplifier gain is represented by $G$ and FBGs's reflectivities are denoted by $R_1$ and $R_2$. The cavity delay in forward and backward directions are represented by $\delta_f$ and $\delta_b$, respectively, and are given by the following expressions:
\begin{equation}
\begin{split}
\label{Cha4_Eq:3}
\delta_f = \frac{2 \pi n_f l_f}{\lambda_0}+\frac{2 \pi n_{t} l_t}{\lambda_0}, \\
\delta_b = \frac{2 \pi n_f l_b}{\lambda_0}+\frac{2 \pi n_{t} l_t}{\lambda_0},
\end{split}
\end{equation}
where $n_f$ and $n_t$ are effective indices of SMF-28 fiber and tapered fiber, respectively. The forward and backward lengths in the cavity are $l_f$ and $l_b$. We assume $n_f$=1.4457 for all our results and we determine $n_t$ via finite element simulations for a tapered fiber present in a medium.
In the wavelength domain, the following expression represents the full width half maximum~(FWHM):
\begin{equation}
\label{Cha4_Eq16}
\Delta\lambda_{1/2} = \frac{\lambda_0^2 \big(\frac{1}{\sqrt{\chi \sqrt{G}}} - \sqrt{\chi \sqrt{G}}\big)}{\pi \big[n_f(l_f +l_b)+2 n_{tap} l_t\big]},
\end{equation}
where $\chi$ represents the overall losses experienced by the cavity and is given by the following expression:
\begin{equation}
\label{Cha4_Eq:6}
\chi = \sqrt{L_t^2 L_{cn}^9 L_{c1}^{1\rightarrow 2} L_{c1}^{2\rightarrow 3} L_{c2}^{1\rightarrow 2} L_{c2}^{2\rightarrow 3} R_1 R_2}.
\end{equation}
Note that $\sqrt{\chi \sqrt{G}}$ has to be less than 1.

\subsection{PS-CRDS sensing principle of the proposed cavity}\label{subsec:PSCRDS}
The proposed setup employs an amplitude modulated laser source to excite the cavity. The light inside the cavity passes through the tapered fiber, optical circulators, and amplifier multiple times before it decays down to zero. The decay time is related to the phase delay between the cavity output and the reference modulating signal according to the following PS-CRDS relationship \cite{berden2009cavity}:
\begin{equation}
\label{Cha4_Eq20b}
\tan \phi = - \omega_m \tau,
\end{equation}
where $\omega_m$ is the modulation frequency. The amplifier gain is used for overcoming background absorption that is usually induced by liquids at 1550~ nm.  The ring down time will increase with an increase in the gain, and hence the phase, $\phi$, will also increase.

A change in the analyte concentration induces a change in the tapered fiber's surrounding refractive index (RI). Therefore, when the concentration increases, the optical mode is pushed out of the tapered fiber and experiences a loss. Simultaneously, the analyte concentration will also impact absorption loss of the propagating mode's evanescent field. These losses decrease the cavity's ring down time, and hence the phase, $\phi$, also reduces. The ring down time of a cavity is also related to its FWHM, as can be easily shown from the energy conservation principles \cite{Jackson_2007},

\begin{equation}
\label{Cha4_Eq20a}
\dfrac{\lambda_0}{\Delta\lambda_{1/2}}=\omega_0 \tau,
\end{equation}
where $\omega_0$ is the optical frequency. Substituting  $\tau$ from equation~\eqref{Cha4_Eq20a} in equation~\eqref{Cha4_Eq20b}, we obtain
\begin{equation}
\label{Cha4_Eq20}
\phi = -\tan^{-1}\bigg(\frac{f_m \lambda_0^2}{(c/n_f)\Delta\lambda_{1/2}}\bigg),
\end{equation}
where $\Delta\lambda_{1/2}$ is given by equation~\eqref{Cha4_Eq16}.
\section{Simulation results}\label{sec:Sim}
For generating simulation results, we assume that the tapered fiber inside the cavity is immersed in aqueous glucose solutions of various concentrations. We use the following expression for determining refractive indices of glucose solutions \cite{yeh2008real}:
\begin{equation}
\label{Cha4_Eq21}
n_{sol} = x_0(\lambda)+y_0(\lambda)C_g,
\end{equation}
where $x_0=1.318$ is the refractive index of DI water at room temperature and at 1550 nm \cite{hale1973optical}, $y_0~=~6.5998~\times~10^{-4}$~(mM)~$^{-1}$ denotes a conversion slope between glucose concentration and solution refractive indices, and $C_g$ stands for glucose concentration measured in mM. These values and equation~\eqref{Cha4_Eq21} indicate that the glucose solution's refractive index increases with an increase in glucose. Therefore,  when we immerse a tapered fiber in the solution, the tapered fiber's leaked evanescent field increases as the glucose concentration increases—consequently, the ring down time and hence phase, $\phi$, decreases.

The tapered fiber diameter also influences the evanescent field strength; the thinner the tapered fiber, the stronger the evanescent field and vice versa.
The fraction of the evanescent field outside the tapered fiber, i.e., the ratio of evanescent field power to the total optical power in a cross-sectional area of the tapered fiber, is given as follows:
\begin{equation}
\label{Cha4_Eq22}
\Gamma = \frac{\int_{r_t}^{\infty}P(r,\phi)rdrd\phi}{\int_0^{\infty}P(r,\phi)rdrd\phi}.
\end{equation}

The sample absorption will also impact the overall losses experienced by the tapered fiber. The absorption losses occur due to the analyte (glucose) as well as the background water. The following expression gives the total absorption coefficient, $\alpha$, of a glucose aqueous solution relative to air \cite{amerov2004molar}:
\begin{equation}
\label{Cha4_Eq23}
\alpha = \Gamma \alpha_w C_w + \Gamma \alpha_g C_g - \Gamma \alpha_w f_w^g C_g,
\end{equation}
where $\alpha_w$ is the absorption coefficient of DI water, $\alpha_g$ is the absorption coefficient of glucose, $C_w$ is the DI water concentration, $C_g$ is the glucose concentration, and $f_w^g$ is the water displacement coefficient of glucose. In equation~\eqref{Cha4_Eq23}, the first and second terms represent absorption due to water and glucose, respectively. The last term deducts a portion of water absorption from that solution part, which is replaced by glucose.

The increase in the glucose concentration of a solution brings two effects in the picture, i.e., a change in the refractive index and a change in the solution's absorption. Furthermore, the refractive index's increase enhances the evanescent field and a corresponding increase in $\Gamma$ of the tapered fiber. The large evanescent field interacts with a large solution portion, and hence, the absorption loss increases, as predicted by equation~\eqref{Cha4_Eq23}. In parallel, an increase in glucose concentration also changes the absorption loss of the solution such that, if $\alpha_g < \alpha_w f_w^g$, there is a decrease in the absorption loss while for $\alpha_g > \alpha_w f_w^g$ the absorption loss increases. The solution's net absorption loss is a combined effect of an increase in the loss due to the evanescent field and a change in the absorption loss due to increased glucose concentration.

\begin{figure}[ht]
\centering
\includegraphics[width=8.5cm]{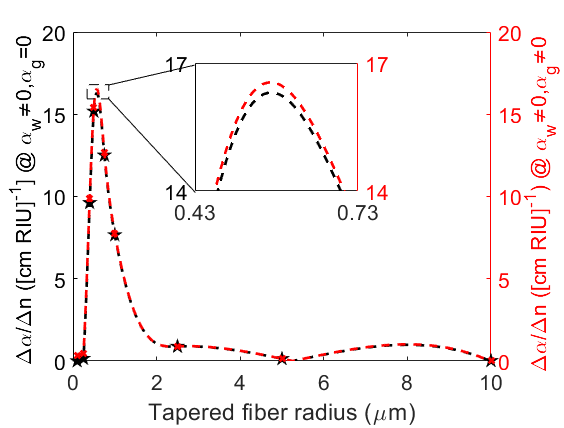}
\caption{The change in tapered fiber absorption as a function of tapered fiber’s radius for given RI changes. Each data point is generated for glucose solutions of concentrations 0-15.5~mM. See the text for details.}
\label{Ch4_Fig_4}
\end{figure}

For the current work, we use~$\alpha_w=8.3633\times 10^{-5} (mM)^{-1} cm^{-1}$ \cite{hale1973optical}, $C_w=55135 mM$, $C_g=(0-15.15) mM$, $\alpha_g=7.23\times 10^{-4} (mM)^{-1} cm^{-1}$\cite{amerov2004molar}, and $f_w^g=6.245$ \cite{amerov2004molar}. We use the finite element method along with equations~\eqref{Cha4_Eq21}-\eqref{Cha4_Eq22} to determine $\Gamma$ for a range of tapered fiber radii immersed in various glucose concentration solutions. Using equation \eqref{Cha4_Eq23}, we then obtain a linear curve of $\alpha$ for each tapered fiber radius as a function of glucose concentrations. Finally, we plot each $\alpha$ curve slope against the tapered fiber's corresponding radius, as shown in Fig.~\ref{Ch4_Fig_4}. Each data point is generated for refractive index changes of $10^{-4}-10^{-2}$ which correspond to 0-15.15~mM glucose solutions. The black plot considers only water absorption loss, i.e., an ideal solution that has the same absorption coefficient as water but induces refractive index changes equivalent to glucose solutions. In contrast, the red plot considers both water and glucose absorption losses for glucose solutions. In the case of glucose solution, $\alpha_g > \alpha_w f_w^g$, which indicates that the addition of glucose will enhance the solution's absorption loss compared to pure DI water, as also evident from Fig.~\ref{Ch4_Fig_4}. These results indicate that the performance of the PS-CRDS sensor will be optimum on using the tapered fiber diameter of $\approx$ 1.1~$\mu m$ for detecting glucose in aqueous solutions. Note that this optimum diameter will vary for different analyte solutions. 

The following expression now represents the loss experienced by the tapered fiber:
\begin{equation}
\label{cha4_24}
L_{t} = e^{-\alpha l_t}.
\end{equation}
We assume $l_t$=2.5~mm, $l_f$=9.24~m, and $l_b$=6.24~m. The overall cavity loss is given by $\chi$ (see equation \eqref{Cha4_Eq:6})and for the constant setup losses we assume, $L_{cn}$=0.15dB, $L_{c1}^{1\rightarrow 2}$=0.88dB, $L_{c1}^{2\rightarrow 3}$=0.79dB, $L_{c2}^{1\rightarrow 2}$=0.57dB, $L_{c1}^{2\rightarrow 3}$=0.6dB, $R_1$=0.86, and $R_2$=0.999. We measure these losses and lengths for our experimental setup in our lab.

\begin{figure}[ht]
\centering
\includegraphics[width=8.5cm]{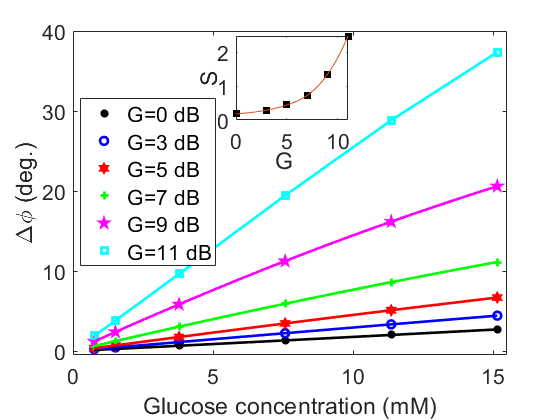}
\caption{Simulated phase shift as a function of glucose concentrations w.r.t pure DI water for 1.1$\mu m$ diameter tapered fiber at various amplifier gains. The inset shows the plot of sensitivity, S, in degrees/mM as a function of amplifier gain, G, in dB. The sensitivity points are calculated by determining the slope of each line.}
\label{Ch4_Fig_9}
\end{figure}

\begin{figure}[ht]
\centering
\includegraphics[width=8.5cm]{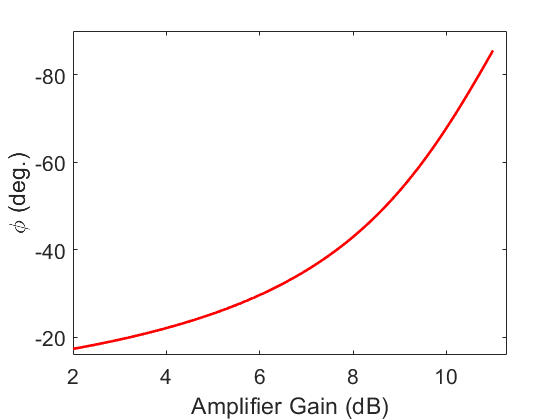}
\caption{Simulated phase shift in water as a function of the amplifier gain.}
\label{Ch4_Fig_5}
\end{figure}

Now using aforementioned simulation results, losses, and equation~\eqref{Cha4_Eq20}, we simulate phase shift, $\phi$, for various glucose concentrations at different amplifier gains and at a modulation frequency of 1MHz. Our simulation results are shown in Fig.~\ref{Ch4_Fig_9}, where we are plotting difference in phase shifts, $\Delta \phi$, between pure DI water and glucose solutions. We then plot each line's slope in Fig.~\ref{Ch4_Fig_9} inset, which shows that sensor's sensitivity increases with the amplifier gain.
We also plot the phase shift for various glucose concentrations against the amplifier gain, as shown in Fig.~\ref{Ch4_Fig_5}. We assume the optimum tapered fiber diameter of 1.1 $\mu m$ in these simulation results. These results indicate that the non-amplified sensitivity of 0.18$^o$/mM or 273$^o$/RIU can be increased to 2.48$^o$/mM or 3765$^o$/RIU by using an active cavity with a gain of 11~dB, i.e., the sensitivity enhancement of around fourteen times is possible.

\section{Experimental Results}\label{sec:Exp}
This section presents our experimental results of sensing glucose in aqueous solutions using our amplified PS-CRDS setup. First, we prepare various concentrations of glucose solutions. These solutions are then injected into a custom-made fluidic cell containing the tapered fiber via a standard syringe. We fabricate the tapered fiber of $\approx$ 7~$\mu m$ diameter and 2.5~mm waist length by pulling SMF-28 fiber and heating it simultaneously with a butane flame. We wash the tapered fiber and the fluidic cell thoroughly with DI water before injecting every glucose sample.

As shown in Fig.~\ref{Ch4_Fig_1}, we build the setup by using Eblana Photonics laser diode~(EP1550-0-NLW-B26-100FM),  Oeland Inc. Canada FBGs with R$_1$=86\%, R$_2$=99.9\% at 1550~nm, lock-in amplifier~(Zurich Instruments MFLI), Mach-Zehnder modulator~(Sumitomo Osaka T.MZH1.5-10PD-ADC), various Thorlabs components, including laser controller~(CLD1015), polarization controller~(FPC032), circulators~(6015-3-APC), silicon amplifier~(S7FC1013S), and photodetector~(DET08CFC/M).

\begin{figure}[ht]
\centering
\includegraphics[width=\linewidth]{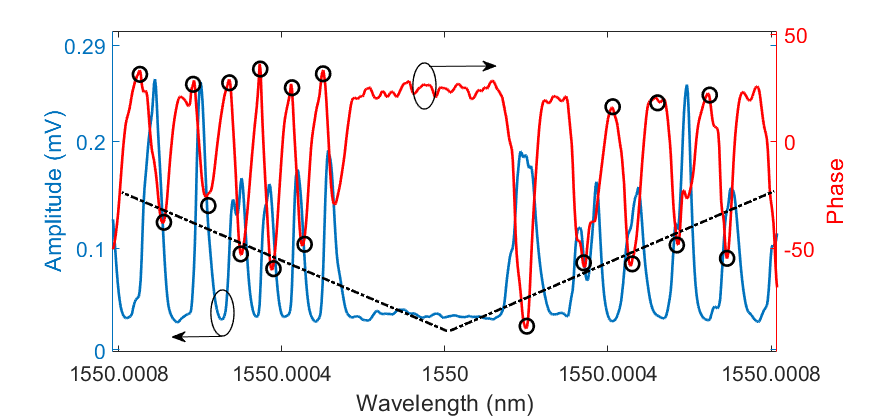}
\caption{Representative amplified PS-CRDS experimental data. The black dotted line is the wavelength sweep signal obtained via the laser current tuning. The blue and red curves are amplitude and phase data, respectively. The black circles on the phase curve are data points for determining the phase for a glucose concentration.}
\label{Ch4_Fig_6}
\end{figure}

The laser is current tuned by a triangular signal of amplitude 50~mV$_{pp}$ and frequency 10 Hz. The current tuning enables the laser wavelength to be swept over a range of $\approx~$23~pm and centered at 1550~nm. Simultaneously, the laser is amplitude modulated at 1~MHz using the Mach-Zhander modulator. A representative amplified PS-CRDS measurement for the active cavity setup is shown in Fig.~\ref{Ch4_Fig_6}. We obtain absolute phase shifts by subtracting the phase minimum and maximum for a resonance peak as shown with black circles in Fig.~\ref{Ch4_Fig_6}. The absolute phases are then averaged over 50-60 phase dips to produce a single-phase measurement for an injected sample. To produce the error bars, we repeat measurements seven times for each sample, i.e., each error bar is produced from the data of 350-420 phase dips.

\begin{figure}[ht]
\centering
\includegraphics[width=8.5cm]{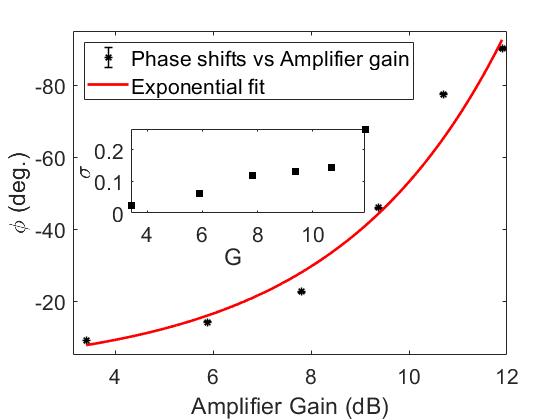}
\caption{Experimentally observed phase shift as a function of the amplifier gain. We use a tapered fiber of diameter $\approx$ 7 $\mu m$ and is placed in DI water. The inset shows the standard deviation, $\sigma$, in degrees of each error bar at the measured amplifier gains, G, in dB.}
\label{Ch4_Fig_7}
\end{figure}

\begin{figure}[ht]
\centering
\includegraphics[width=8.5cm]{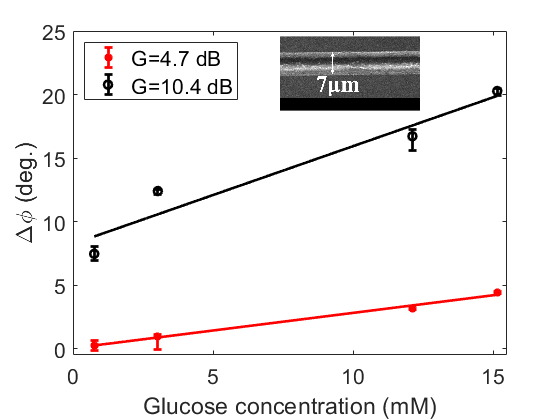}
\caption{Experimental results for PS-CRDS phase shift as a function of glucose concentrations. The inset shows the SEM image of the thinnest part of the experimental tapered fiber.}
\label{Ch4_Fig_8}
\end{figure}

The amplifier gain impact on the PS-CRDS phase measurements is shown in Fig.~\ref{Ch4_Fig_7}. The increase in the amplifier gain increases the cavity phase delay as the light will stay longer inside the cavity, as shown in Fig.~\ref{Ch4_Fig_7}.
The measured phase shift, with respect to pure DI water, for various glucose concentrations is shown in Fig.~\ref{Ch4_Fig_8}. From the experimental data, we obtain the sensitivity and detection limit of 0.768$~^o$/mM (1164$~^o$/RIU) and 0.75 mM ( 4.5~$\times$~10$^{-4}$~RIU), respectively.

\section{Conclusions}\label{sec:concl}
Our experimental results are clearly in agreement with the theoretical ones, where we show that phase shift and hence the cavity ring down time in aqueous solutions increases exponentially as a function of the amplifier gain~(see Fig.~\ref{Ch4_Fig_5} and Fig.~\ref{Ch4_Fig_7}). We also show that the sensitivity of an active PS-CRDS sensor increases with the amplifier gain~(see  Fig.~\ref{Ch4_Fig_9} and Fig.~\ref{Ch4_Fig_8} ). These results show that our demonstrated active sensor reduces the background absorption of liquids at 1550~nm significantly, which is a major hindrance for developing biosensors or chemical identification sensors at 1550~nm.

Our novel setup also allows light to transverse through the tapered fiber twice in one round trip~(see Fig.~\ref{fig:loopvslinear}). It hence potentially offers superior performance compared to conventional fiber loops. As a comparison, our current experimental results show an approximately five times higher detection limit compared to a previous work involving CRDS with active fiber loops at 1550~nm \cite{sharma2018comparison}. However, it should be noted that the numbers reported in this work are not the absolute limits of the demonstrated sensor as currently in our lab, we can not fabricate a tapered fiber of the theoretical optimum diameter of 1.1~$\mu m$ for glucose detection in aqueous solutions. We also have not employed any temperature controllers for the FBGs and fluidic cell to mitigate the ambient temperature variations during experiments. Furthermore, we assume that the tapered fiber comprises only a uniform thinner portion in simulation results. However, there are always transition regions where SMF-28 fiber diameter is changing from 125$~\mu m$ to a minimum diameter. We will consider the points mentioned above in future work towards a biosensing application with active fiber cavities. Another future direction is to extend our work towards active ring resonators for liquid sensing applications at 1550~nm.

In summary, we demonstrate a linear and active fiber cavity PS-CRDS sensor with a novel design that is expected to find a wide range of sensing applications in food security and biological media.

\bibliographystyle{unsrt}
\bibliography{References}

\end{document}